\documentclass[aip, jap, amsmath,amssymb,citeautoscript,
 reprint,%
]{revtex4-1}
\usepackage{etoolbox}
\usepackage{graphicx}
\usepackage{bm}

\newcommand{\ud}{\,{\mathrm d}}

\newcommand{\uz}{z}
\newcommand{\uzc}{\overline{z}}
\newcommand{\uw}{w}
\newcommand{\uwc}{\overline{w}}
\newcommand{\uf}{f}
\newcommand{\ufc}{\overline{f}}

\newcommand{\uiiint}{\int\!\!\!\int\!\!\!\int}

\newtoggle{preprint}
\toggletrue{preprint}
\newtoggle{submission}
\toggletrue{submission}
\iftoggle{preprint}{%
  \allowdisplaybreaks %
}{%
} %
\begin{document}

\title{Vortex precession frequency and its amplitude-dependent shift
  in cylindrical nanomagnets}

\author{Konstantin L. Metlov}
\email{metlov@fti.dn.ua}
\affiliation{Donetsk Institute for Physics and Technology NAS, Donetsk, Ukraine 83114}
\date{\today}
\begin{abstract}
Frequency of free magnetic vortex precession in circular soft ferromagnetic
nano-cylinders (magnetic dots) of various sizes is an important parameter,
used in design of spintronic devices (such as spin-torque microwave 
nano-oscillators) and characterization of magnetic nanostructures. 
Here, using a recently developed
collective-variable approach to non-linear dynamics of magnetic textures
in planar nano-magnets, this frequency and its amplitude-dependent
shift are computed analytically and plotted for the full range of cylinder
geometries. The frequency shift is positive in large planar dots, but becomes
negative in smaller and more elongated ones. At certain dot dimensions a
zero frequency shift is realized, which can be important for enhancing
frequency stability of magnetic nano-oscillators.
\end{abstract}

\pacs{75.78.Fg, 75.70.Kw, 75.75.Jn} \keywords{magnetization dynamics,
  magnetic nano-dots, magnetic vortex, anharmonism} \maketitle

Cylindrical magnetic nano-pillars with magnetic vortex are an important
component of emerging spintronic applications, such as magnetic random
access memory (MRAM)\cite{VanWaeyenberge2006,Cowburn_NatM_2007} with 
ultra-fast core switching\cite{LGLK07,HGFS07} or spin-torque microwave
nano-oscillators\cite{Pribiag_Vortex_NatPhys07}. The problem
of non-linear magnetization dynamics in such systems poses a 
significant fundamental challenge. Only recently the amplitude-dependent 
frequency shift of vortex precession was probed experimentally 
in Ref.~\onlinecite{SPLNKMAMG13}. This paper fully explores the 
problem of non-linear vortex precession in a circular magnetic nano-cylinder 
and describes the measured frequency shift\cite{SPLNKMAMG13} quantitatively using
a recently developed collective-variable approach to 
magnetization dynamics\cite{M13.dynamics}.

Let us briefly remind the essentials of the approach. 
It is based on the Lagrangian formalism and allows to obtain the equations 
of motion for a magnetization texture, parametrized via a number of collective
coordinates. A parametrization for a displaced magnetic vortex
was developed earlier\cite{M01_solitons2}, which is a particular case of a more
general family of trial functions\cite{M10} and had already been successfully
used in computing the vortex precession frequency in {\em large} planar
magnetic dots\cite{GHKDB06,M13.dynamics}. It is given by the following
analytic function of complex variable
\begin{equation}
  \label{eq:noside}
  f(z)=\imath \frac{z - (A + \overline{A} z^2)}{r_V},
\end{equation}
where $\imath=\sqrt{-1}$, $z=X+\imath Y$; $X$, $Y$ and $Z$ are
the Cartesian coordinates with the origin at the center of the cylinder 
($Z$ is parallel to its axis); 
$r_V=R_V/R$ is dimensionless vortex core radius and 
$A=A(t)=a_X(t)+\imath a_Y(t)$, $|A|<1/2$
is a complex pair of collective coordinates. Position of the vortex center
$f(z_C)=0$ is 
$z_C=(1-\sqrt{1-4 A \overline{A}})/(2 \overline{A})$.
The centered vortex $z_C=0$ corresponds to $A=0$. 
An associated complex function\cite{M10}
\begin{equation}
  \label{eq:sol_SM2}
  \uw(\uz,\uzc)=\left\{
    \begin{array}{ll}
      \uf(\uz) & |\uf(\uz)| \leq 1 \\
      \uf(\uz)/\sqrt{\uf(\uz) \ufc(\uzc)} & |\uf(\uz)| >1
    \end{array}
    \right. ,
\end{equation}
where the region $|\uf(\uz)| \leq 1$ corresponds
to the vortex core (soliton) allows to recover 
the magnetization vectors $\vec{m}=\vec{M}/M_S$
via the stereographic projection $m_X+\imath m_Y=2 \uw/(1+\uw\uwc)$,
$m_Z=\pm (1-\uw\uwc)/(1+\uw\uwc)$,
which ensures that $|\vec{m}|=1$. The $\pm$ sign in $m_Z$ corresponds to
different polarizations of the vortex core: $m_Z=\pm 1$ at the
vortex center. The function $\uw(\uz,\uzc)$ is 
analytic inside the vortex core and non-analytic outside.

Dynamics of the magnetization texture (\ref{eq:noside}) can be computed from
its Lagrangian (normalized here by $\mu_0 M_S^2 \pi L_Z R^2$, where
$\mu_0$ is permeability of vacuum, $L_Z$ is cylinder's thickness,
$R$ is its radius). Assuming the absence of the external driving forces
and smallness of the vortex core center 
displacements $|A|\ll 1$, the Lagrangian has the form\cite{M13.dynamics}
\begin{eqnarray}
  {\cal L} & = &
  \pm ( \kappa_2 + \kappa_4 (a_X^2\!+\!a_Y^2))\left(a_X\dot{a}_Y\!-\!a_Y\dot{a}_X\right) \nonumber \\
  & & - k_2\left(a_X^2\!+\!a_Y^2\right)   
  - k_4\left(a_X^2\!+\!a_Y^2\right)^2 + \ldots ,
  \label{eq:Lnoside4}  
\end{eqnarray}
where the dot over variable denotes 
time derivative,$\kappa_2=(1+r_V^4(4 \log\!2 - 3))/(\gamma \mu_0 M_S)$ and 
$\kappa_4=2-r_V^2(23+r_V^2((6061 - 6397 r_V^2)/8 -
1152(1-r_V^2)\log 2))/(\gamma \mu_0 M_S)$ have units of seconds,
$\gamma\simeq 1.76 \cdot 10^{11}\mathrm{rad}/(\mathrm{s}\,\mathrm{T})$
is gyromagnetic ratio; $k_2$ and $k_4$ are dimensionless potential
energy expansion coefficients. The first line in (\ref{eq:Lnoside4}) 
is the kinetic energy and the second line is the negative
potential energy of the vortex.

The equations of motion are the Euler-Lagrange equations extremizing the 
Lagrangian (\ref{eq:Lnoside4}), they are
\begin{eqnarray}
 (\kappa_2+2\kappa_4(a_X^2+a_Y^2))\dot{a}_X\pm
 (k_2+2 k_4 (a_X^2+a_Y^2))a_Y & = & 0 \\
 (\kappa_2+2\kappa_4(a_X^2+a_Y^2))\dot{a}_Y\mp
 (k_2+2 k_4 (a_X^2+a_Y^2))a_X & = & 0 .
\end{eqnarray}
For the initial conditions $a_X(0)=a_0$, $a_Y(y)=0$ these equations are
solved by\cite{M13.dynamics}
\begin{equation}
a_X(t) = a_0 \cos(\omega t), \,\, 
a_Y(t) = \pm a_0 \sin(\omega t) \label{eq:nocharges2sol}
\end{equation}
with
\begin{eqnarray}
  \omega & = &  \frac{k_2 + 2 a_0^2 k_4}{\kappa_2 + 2 a_0^2 \kappa_4}\simeq
  \omega_0 + 2 \alpha a_0^2 + \ldots \label{eq:freq4} \\
  \omega_0 & = & k_2/\kappa_2, \label{eq:omega0}, \,\, 
  \alpha = \frac{k_4 \kappa_2 - k_2 \kappa_4}{\kappa_2^2} ,\label{eq:shift}
\end{eqnarray}
where $\alpha$ is the coefficient, relating the frequency of vortex rotation
to its amplitude. These formulas are simple, but the dependence of the 
potential energy expansion coefficients $k_2$ and $k_4$ on cylinder 
dimensions can be rather complex. Evaluation and analysis of this 
dependence is the main subject of this work.

For the cylinder made of soft ferromagnetic material the main contributions
to the potential energy come from the exchange and dipolar interactions.
The exchange energy is simplest to evaluate, it is
\begin{equation}
E_{EX}=\frac{C M_S}{2} \uiiint_V 
\sum\limits_{i=X,Y,Z}(\vec{\nabla} m_i(\vec{r}))^2 \ud^3\vec{r},
\end{equation}
where $C=2A$ is the exchange stiffness and 
$\vec{\nabla}=\{\partial/\partial X, \partial/\partial Y, \partial/\partial Z\}$.
For the magnetization distribution (\ref{eq:noside}) the integration can be
efficiently carried out using the residue theorem\cite{M12.core} even without
assuming smallness of the vortex center displacements,  which yields\cite{MG02_JEMS}
\begin{eqnarray}
 e_{EX}&=&\frac{E_{EX}}{\mu_0 M_S^2 \pi L_Z R^2} =
 \frac{1}{\rho^2}
 \left(2-\log \frac{2 r_V}{1+\sqrt{1-4 |A|^2}}\right) \nonumber \\
 \label{eq:exchange}
 & = & \frac{2 - \log r_V}{\rho^2}-
 \frac{|A|^2}{\rho^2}-\frac{3 |A|^4}{2 \rho^2}-\ldots
\end{eqnarray}
where $\rho=R/L_E$, $L_E=\sqrt{C/(\mu_0 M_S^2)}$ is the exchange length
of cylinder's material. The exchange interaction is pushing the vortex
out of the cylinder, as manifested by the negative signs
before the powers of $|A|$.

Computation of the dipolar energy is more involved. To make
it simpler, let us use the magnetostatic approximation by neglecting
the time derivatives in Maxwell's equations and assuming that there are
no macroscopic currents in the magnet. This approximation is justified
{\it a posteriori} by noting that at characteristic GHz frequencies, obtained
as the result of this computation, there is plenty of time for electromagnetic
waves to propagate many times throughout the entire sub-micron ferromagnetic
cylinder and, thus, for demagnetizing field to be velocity-independent. 
The magnetostatic approximation allows to introduce the {\em scalar} potential
for demagnetizing field and
express the magnetostatic energy as the interaction energy of magnetic
charges with the density $-M_S (\vec{\nabla}\cdot\vec{m})$. Further, noting that at the
boundary of the magnetic material the divergence is equal to the normal component
of magnetization vector, magnetic charges can be subdivided into the
surface charges (located
at the faces of the cylinder in this case) and the volume charges 
(they become non-zero when vortex is displaced from the face center, $|A|>0$).
There are no surface charges on cylinder's side and also, due to symmetry 
(charges on the different faces have the opposite sign), there is no
interaction between the face and volume magnetic charges.

The energy of face charges (in the same normalization by 
$\mu_0 M_S^2 \pi L_Z R^2$) on both faces and their mutual
interaction can be written as $e_\mathrm{f}= U(0) - U(g)$ with
\begin{eqnarray}
 U(h) & = &
 \int\limits_0^{2\pi} 
 \int\limits_0^{\theta(\varphi_1)} 
 \int\limits_0^{2\pi} 
 \int\limits_0^{\theta(\varphi_2)}\!\!\!\!
 u(h)
 \ud\varphi_1
 \ud r_1
 \ud\varphi_2
 \ud r_2
 \\
 u(h)&=&
 \frac{m_Z(r_1, \varphi_1)m_Z(r_2,\varphi_2) r_1 r_2   }
{\sqrt{r_1^2\!+\!r_2^2\!-\!2 r_1 r_2 \cos (\varphi_1\!-\!\varphi_2)\! +\! h^2}},
\end{eqnarray}
where $g=L_Z/R$ is cylinder's aspect ratio and $r=\theta(\varphi)$ is the
equation of vortex core boundary in polar coordinate system $r$,
$\varphi$ centered in the vortex center $z_C$, and with the origin of polar
angle chosen to coincide with the direction of the complex phase of $A$.
When $|A|=0$ the vortex core boundary is a circle of
radius $r_V$, for larger $|A|$ it becomes deformed. This
deformation must necessarily be taken into account to obtain the correct 
values of magnetostatic energy. The equation for
vortex core boundary is 
$f(r \exp(\imath \varphi) - z_C)=1$. It has no explicit
analytical solutions for $r$, but can be solved by 
$r=\theta(\varphi)$ in the form of Taylor series:
\begin{eqnarray}
 \frac{\theta(\varphi)}{r_V} & =  & 1 + b_1 |A| + b_2 |A|^2 + b_3 |A|^3 
 + b_4 |A|^4 + \ldots ,
\end{eqnarray}
with $b_1 = r_V \cos \varphi$, $b_2 = (5 b_1^2 + 4 -r_V^2)/2$,
$b_3 = b_1 (8 b_1^2 + 6 - 3 r_V^2))$, 
$b_4 = (48+200b_1^2+231b_1^4-2(20+63b_1^2)r_V^2+7r_V^4)/8$.
The density of the face charges is 
$m_Z(r,\varphi) = (r_V^2 - \beta^2)/(r_V^2 + \beta^2)$, 
with
\begin{eqnarray}
 \beta^2 & = & r^2(1-|A|^2(4-r^2)) - 
 2 |A| r^3 \sqrt{1-4|A|^2}\cos \varphi . \label{eq:theta}
\end{eqnarray}
After the Taylor expansion  $m_Z$ also contains both even and
odd powers of $|A|$ and also depends on $\varphi$. 
By carefully evaluating all the terms it is possible to obtain
the following expansion for the magnetostatic energy of face charges
\begin{eqnarray}
 \label{eq:eface}
 e_\mathrm{f}&=& \frac{r_V^3}{g}(W_{00}(\frac{g}{r_V})+
 |A|^2(W_{20}(\frac{g}{r_V})+r_V^2 W_{22}(\frac{g}{r_V}))+ \nonumber \\
 & & |A|^4(W_{40}(\frac{g}{r_V})+r_V^2 W_{42}(\frac{g}{r_V})+
 r_V^4 W_{44}(\frac{g}{r_V}))),
\end{eqnarray}
where the magnetostatic functions $W_i$ are defined in the
\iftoggle{preprint}{%
  Appendix A.
}{%
  Supplemental Material\cite{suppfrequency}.
} The function $W_{00}(h)$ coincides with the magnetostatic function
derived by Usov and Peschany\cite{UP93} for centered vortex. 

The density of the volume charges is proportional to the negative 
divergence of the magnetization
vector field $\Omega = - (\vec{\nabla} \cdot \vec{m})$, having different expressions
inside and outside the vortex core
\begin{equation}
 \Omega(r,\varphi) = \left\{
 \begin{array}{ll}
  \frac
 {8 |A| r r_V^3 \sin\varphi}
 {(r_V^2 + \beta^2)^2} & 0 \leq r<\theta(\varphi) \\
  \frac
 {2 |A| r \sin\varphi}
 {\beta} & \theta(\varphi)<r\leq\Theta(\varphi)
 \end{array}
 \right.,
\end{equation}
where $\beta$ is defined by (\ref{eq:theta}) and 
$r=\Theta(\varphi)$ is the equation for cylinder boundary
in the polar coordinate system, centered in the vortex center $z_C$
\begin{eqnarray}
 \Theta(\varphi)& = & 1 - c_1 |A| - c_2 |A|^2 - c_3 |A|^3 - c_4 |A|^4 + \ldots,
\end{eqnarray}
with $c_1=\cos\varphi$, $c_2=(1/2)\sin^2\varphi$, $c_3=\cos\varphi$,
$c_4=(17 - \cos(2 \varphi)\sin^2\varphi)/16$.
The energy of the volume charges is then
\begin{eqnarray}
 \label{eq:evolint}
 e_\mathrm{v}&=&
 \frac{1}{2 g}\!
 \int\limits_0^{g} \!\!
 \ud z_1 \!\!
 \int\limits_0^{2\pi} \!\!
 \ud\varphi_1\!\!\!\!
 \int\limits_0^{\Theta(\varphi_1)}\!\!\!\!
 \ud r_1\!\!
 \int\limits_0^{g} \!\!
 \ud z_2 \!\!
 \int\limits_0^{2\pi}\!\!
 \ud\varphi_2\!\!\!\!
 \int\limits_0^{\Theta(\varphi_2)}\!\!\!\!
 \ud r_2
 \,w \\
 w&=&
 \frac{\Omega(r_1, \varphi_1)\Omega(r_2,\varphi_2) r_1 r_2   }
{\sqrt{r_1^2\!+\!r_2^2\!-\!2 r_1 r_2 \cos (\varphi_1\!-\!\varphi_2)\! +\! (z_1-z_2)^2}},
\end{eqnarray}
where the factor $1/2$ is due to the fact that each pair of interacting
charges is summed twice during the double volume integration. Expanding
it in powers of $|A|$ gives
\begin{equation}
 \label{eq:evolume}
 e_\mathrm{v} =  |A|^2 V_2(r_V, g) + |A|^4 V_4(r_V, g)
\end{equation}
with the magnetostatic functions $V_i$, defined in the
\iftoggle{preprint}{%
  Appendix B.
}{%
  Supplemental Material\cite{suppfrequency}.
} While the
volume magnetostatic functions are not much more complex, compared to the
surface ones, they do not admit factorization of their dependence on
$r_V$ and $g$.

Collecting the terms in (\ref{eq:exchange}), (\ref{eq:eface}),
and (\ref{eq:evolume}) allows to recover the potential energy expansion coefficients
\begin{eqnarray}
 k_2 & = & -\frac{1}{\rho^2} + V_2+ 
 \frac{r_V^3}{g} \left( W_{20}+
 r_V^2 W_{22} \right) \label{eq:k2} \\
 k_4 & = & -\frac{3}{2\rho^2} + V_4+ 
 \frac{r_V^3}{g} \left( W_{40}+
 r_V^2 W_{42} + r_V^4 W_{44} \right), \label{eq:k4}
\end{eqnarray}
where the arguments of the magnetostatic functions have been omitted for 
brevity and are the same as in (\ref{eq:eface}),(\ref{eq:evolume}).
Zero-order terms lead to the equation for the equilibrium vortex
core radius\cite{UP93, M12.core} $\rho_V= R_V/L_E$ 
\begin{equation}
 \label{eq:rhov}
 -\frac{1}{\rho_V} + 
 \frac{3 \rho_V^2 W_{00}(\zeta/\rho_V)}{\zeta} -
 \rho_V  W_{00}'(\zeta/\rho_V) = 0 ,
\end{equation}
which depends only on cylinder's thickness $\zeta=L_Z/L_E$.

The formulae (\ref{eq:k2}), (\ref{eq:k2}), (\ref{eq:rhov}), and the 
definitions of magnetostatic functions contain all the 
ingredients necessary to compute the vortex precession frequency $\omega_0$ and
its shift $\alpha$ from (\ref{eq:omega0}). All these
formulas are attached in the form of {\tt MATHEMATICA} script {\tt frequency.m}
as a Supplemental Material\cite{suppfrequency}.
The computed frequencies and shifts are plotted 
in Figures 1 and 2 respectively. The selection of the coordinate system and
range of parameters on these figures was done with intention to show the dependencies
in full for as large set of dot geometries as possible.
Consequently, some of the points in these
figures may correspond to the dot dimensions beyond
the applicability of the present model. At some of them the magnetic vortex might not
be the ground state of the dot (see e.g. phase diagram in Ref.~\onlinecite{ML08}), 
or may be unstable\cite{UP94,MG02_JEMS} at some of the larger included thicknesses
the vortex might develop a 3-d structure, or its motion will cause the spin-wave
generation, making the high frequencies (on the left end of curves in Fig.1) unreachable.
Also, plotting dependencies in full, has the
drawback of blurring their fine details. To make plots with different
scales and axes, please use the attached {\tt MATHEMATICA} script {\tt frequency.m} to
compute the frequencies, shifts and potential energy expansion coefficients for an
arbitrary cylinder geometry and material parameters.
\begin{figure}
\label{fig:frequency}
\iftoggle{submission}{%
  \includegraphics[scale=0.43]{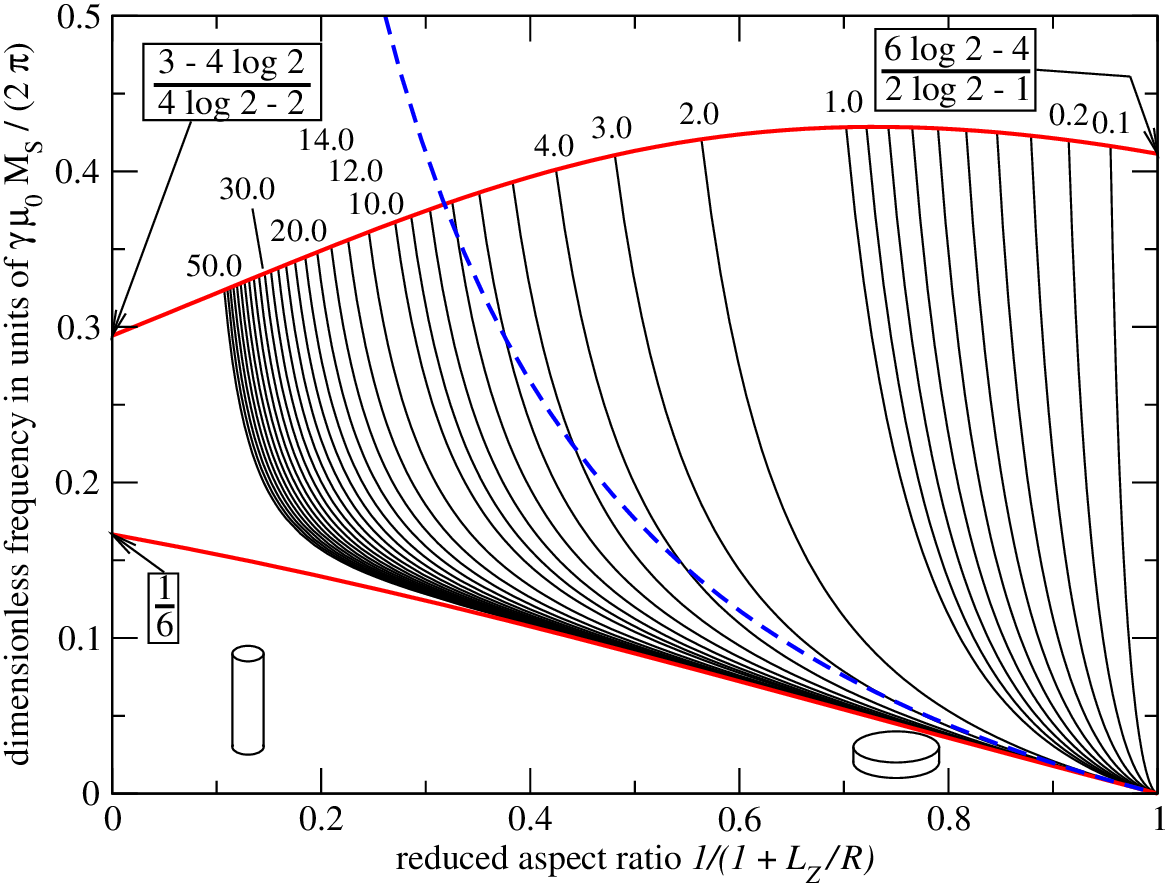} %
}{%
  \includegraphics[scale=0.43]{figures/frequencies}          %
} %
\caption{Frequencies of free magnetic vortex precession in circular
cylinder of radius $R$ and thickness $L_Z$ made of soft ferromagnetic
material with saturation magnetization $M_S$ and the exchange length $L_E$.
The horizontal axis describes the cylinder's aspect ratio $g=L_Z/R$ with
non-linear scale to cover all of its possible values. Thin solid lines
correspond to different cylinder thicknesses, measured in units of $L_E$. 
The lines are sampled with several different equal steps in between the
specified numerical labels. The lower bounding thick solid line shows
the limit of large cylinders with negligible core, the upper thick solid line
is for the cylinder radius exactly equal to the vortex core radius.
The dashed line is the
first order series expansion of the lower bounding line at $g=0$. Formulas
show the exact analytical expressions for frequencies at certain limiting
geometries. Cylinder sketches illustrate shapes of cylinders
at different ends of the horizontal axis.}
\end{figure} 

There are several lessons, which can be learned from these graphs. First, the
vortex precession frequency increases with decrease of the dot radius and 
increase of its thickness. The maximum is achieved when the particle radius
coincides with the vortex
core radius (the model is valid only when the vortex core fits completely
inside the cylinder). It can be expected that due to over-confinement,
resulting from the complete absence of the side charges, 
the vortex precession frequency for dots approaching 
the vortex core size is overestimated and can be considered an upper bound. 
Another computation, based on generalization of the trial function to
relax the side boundary conditions\cite{MG04,M10} may yield a better precision 
in the high-frequency region. For the low 
frequency region (corresponding to large dots) there is an 
established analytical approximation\cite{GHKDB06, M13.dynamics}, 
based on neglecting the vortex core (assuming it is much smaller 
than the particle size) $\omega_0 = \gamma \mu_0 M_S V_2(0,g)$ or
\begin{eqnarray}
  \nu_0 & = & \frac{\omega_0}{2\pi}=\frac{\gamma \mu_0 M_S}{\pi} \!\!\!
  \int\limits_0^\infty
  \frac{f_\mathrm{MS}(k g)}{k}\!
  \left[
    \int\limits_0^1 \! r J_1(k r) \ud r
  \right]^2 \!\!\!\!\!\ud k, \label{eq:nulargefull}\\
  \nu_0 & \simeq & \gamma \mu_0 M_S g \frac{2(2G-1)}{6\pi^2},
  \,\,\,\,\,\,\, g \ll 1,
  \label{eq:freqasympt}
\end{eqnarray}
where $f_\mathrm{MS}(x)=1-(1-e^{- x})/x$ and
$G\simeq0.915\,966$ is Catalan's constant. The lower solid bounding line 
in Fig.~1 is exactly the Eq.~\ref{eq:nulargefull}. By noticing that it
is very close to a straight line in the chosen coordinate system
it is possible to obtain the following simple approximate 
analytical expression for the vortex precession frequency
\begin{equation}
\label{eq:nuapprox}
\nu_0 \approx  \frac{\gamma \mu_0 M_S}{12\pi}\left(
1 - \frac{1}{1+g}
\right).
\end{equation}
Unlike (\ref{eq:freqasympt}), shown in Fig.~1 by 
the dashed line, it does not contain the assumption of $g \ll 1$ and
covers both flat and elongated cylinders. Compared to the full 
integral (\ref{eq:nulargefull}), it has the largest error of 
7.91\% at $g \approx 0.72$ or $1/(1+0.72)\approx 0.58$
decreasing for smaller and larger $g$. 

\begin{figure}
\label{fig:anharmonim}
\iftoggle{submission}{%
  \includegraphics[scale=0.43]{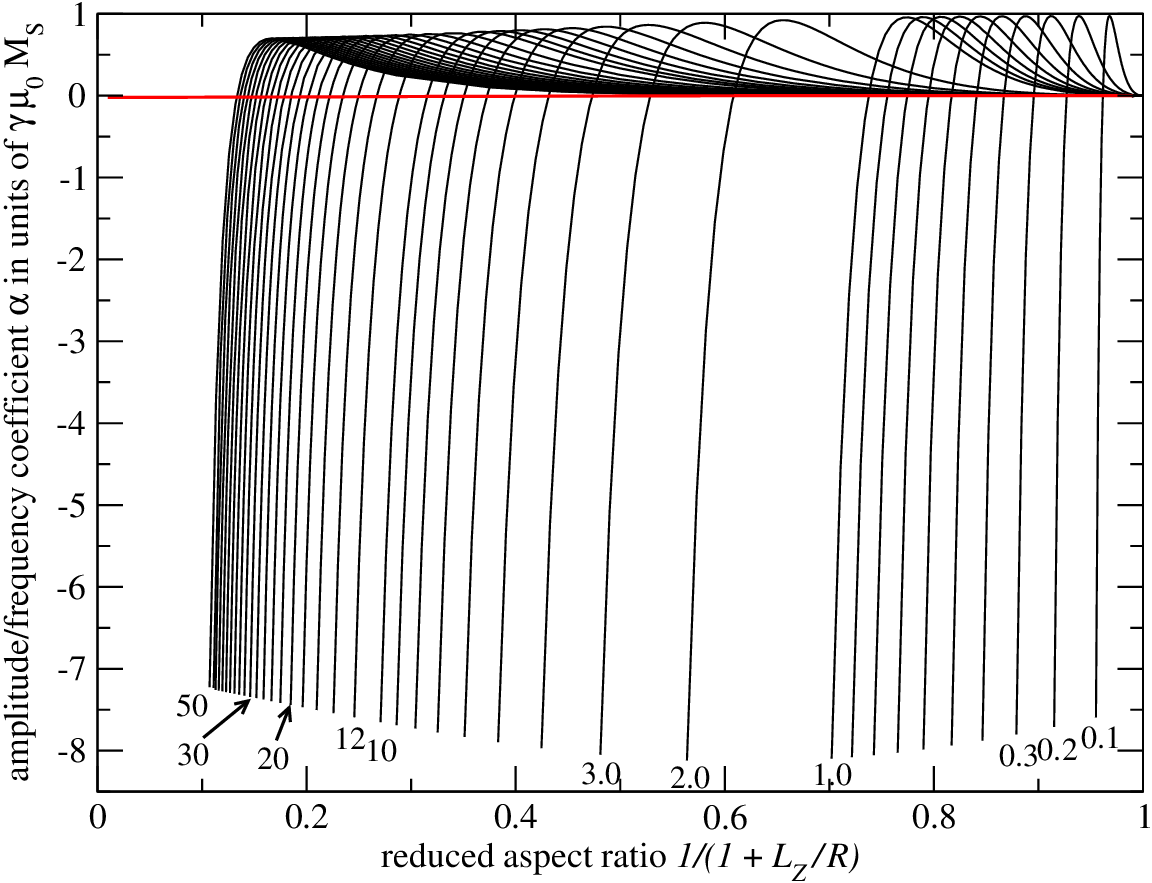} %
}{%
  \includegraphics[scale=0.43]{figures/anharmonism_halved}   %
} %
\caption{Vortex precession frequency shift $\alpha$ at different cylinder
geometries. The horizontal axis, parameters and thicknesses, corresponding to each of
the line of the family, are the same as in Fig.~1.}
\end{figure}
The vortex frequency shift graph in Fig.~2 also covers the full range of
cylinder geometries. Looking at (\ref{eq:shift}) it can be {\it a priori} 
expected that the frequency shift may become negative. This is due to the
negative term, proportional to $\kappa_4$. 
This term describes the change
of the vortex kinetic energy due to its shape deformation, which is missing in other
computations,  based on solution of Thiele 
equation\cite{GHKDB06,IZ07,GHC10,DKBGCF12,SPLNKMAMG13}. The Figure 2 confirms
that it is indeed
possible to have both positive (in cylinders with larger radii)
and negative (in cylinders with smaller radii) frequency shifts. At a
certain cylinder radius the frequency shift turns to zero. This
feature can be useful for achieving high frequency stability of magnetic
vortex precession. The shift is bounded from above by 
$\approx \gamma \mu_0 M_S$, which is achievable in thin dots (about 
$L_E$ in thickness). For thicker dots the maximum frequency shift is 
smaller. 

The dimensionless frequency shift $\lambda$ defined in Ref.~\onlinecite{SPLNKMAMG13} can be 
expressed as
\begin{equation}
 \label{eq:shiftlambda}
 \lambda = \frac{\alpha}{\omega_0} = 
 \frac{k_4 \kappa_2 - k_2 \kappa_4}{k_2 \kappa_2}.
\end{equation}
There is no additional factor despite the fact that $\alpha$ relates 
the frequency shift to $|A|^2$, while $\lambda$ relates it to $|z_C|^2$.
This is because to the first order the quantities $A$ and $z_C$ coincide, 
which is sufficient to obtain the same series expansion of (\ref{eq:freq4}).
The experimental value of  $\lambda$ was determined for a single
FeV dot in Ref.~\onlinecite{SPLNKMAMG13}. Assuming the exchange length $6nm$
for their FeV crystal and the dot radius of $250nm$,
which is the half of the minor axis of the studied elliptical dot that
can be measured from the right half of the inset in their Fig.1, the 
Eq.~\ref{eq:shiftlambda} gives the relative frequency shift
$\lambda=0.38$, which is in agreement with the measured $\lambda=0.5\pm30\%$.

Concluding, the series expansion of the magnetostatic
energy of a cylindrical magnetic dot with a vortex is evaluated up to
the fourth power in the vortex core center displacement, based on the model
with no side magnetic charges (\ref{eq:noside}). It allows to plot
the vortex precession frequency (Fig.~1) and its amplitude-dependent
shift (Fig.~2) for the full range of cylinder geometries. A simple 
algebraic expression for the vortex precession frequency (\ref{eq:nuapprox})
is proposed, which covers a wider set of cylinder geometries, compared to
a similar relation\cite{GHKDB06,M13.dynamics} (\ref{eq:freqasympt}).
It is shown that the frequency shift can be positive and negative for 
certain dot geometries. For every cylinder thickness there is a cylinder radius
for which the amplitude-dependent frequency shift is zero. The value of the
shift is in agreement with the experimental data of 
Ref.~\onlinecite{SPLNKMAMG13}. 

\iftoggle{preprint}{%
\appendix

\section{Magnetostatic functions for face charges}
When expanding the magnetostatic energy of face charges in powers of $|A|$
it is necessary to take into account that both integrand and integration 
limits depend on $|A|$ and to make proper use of the formula for
taking the derivative of an integral, dependent on a parameter:
\begin{eqnarray}
 \frac{\partial}{\partial p} \!\!\int\limits_{l_1(p)}^{l_2(p)} \!\!f(p, x) \ud x & = &
 \!\!\int\limits_{l_1(p)}^{l_2(p)} \!\!\frac{\partial f(p, x)}{\partial p} \ud x + 
 f(p,l_2(p))l_2'(p) - \nonumber \\
 & &  f(p,l_1(p))l_1'(p), \label{eq:integralderiv}
\end{eqnarray}
where prime denotes the derivative. To factor the remaining angular 
and radial integrals,
let us, as usual,
 represent the inverse square root via the Lipshitz integral
\begin{equation}
  \label{eq:lipshitz}
  \frac{1}{\sqrt{b^2+(z_1-z_2)^2}}=\int_0^\infty e^{-k |z_1-z_2|} J_0(k b) \ud k
\end{equation}
and use the Bessel's summation theorem for representing
\begin{eqnarray}
& & J_0(k \sqrt{r_1^2+r_2^2 - 2 r_1 r_2 \cos (\varphi_1-\varphi_2)}) = \nonumber \\
\label{eq:bessel}& & \sum\limits_{\mu=-\infty}^{\infty} J_\mu(k r_1)J_\mu(k r_2)e^{\imath \mu (\varphi_1-\varphi_2)}.
\end{eqnarray}
The magnetostatic energy of face charges is then expressed via the 
magnetostatic function
\begin{eqnarray}
U(h) & = & \sum_{\mu=-\infty}^{\infty}\int_0^\infty e^{- k h} (i(\mu,k))^2 \ud k \\
i(\mu,k) & = & \int_0^{2 \pi} \ud \varphi \int_0^{\theta(\varphi)} \!\!\ud r\,\,
           m_z(r, \varphi) r J_n(k r) e^{\imath \mu \varphi}
\end{eqnarray}
Expanding this expresion in powers of $|A|$ and taking the angular
integrals yields
\begin{eqnarray}
 \label{eq:efaceapp}
 e_\mathrm{f}&=& \frac{r_V^3}{g}(W_{00}(g/r_V)+
 |A|^2(W_{20}(g/r_V)+r_V^2 W_{22}(g/r_V))+ \nonumber \\
 & & |A|^4(W_{40}(g/r_V)+r_V^2 W_{42}(g/r_V)+r_V^4 W_{44}(g/r_V))),
\end{eqnarray}
with magnetostatic functions $W_i(h)=U_i(0)-U_i(h)$ and
\begin{eqnarray}
 i_{0}(k) & = & \int_0^1 \frac{1-r^2}{1+r^2} J_0(k r) r \ud r \\
 U_{00}(h) & = & \int_0^\infty e^{-k h} (i_{0}(k))^2 \ud k \\
 i_{1}(k) & = & \int_0^1 \frac{r^3}{(1+r^2)^2} J_0(k r) \ud r \\
 U_{20}(h) & = & 16 \int_0^\infty e^{-k h} i_{0}(k) i_{1}(k) \ud k \\
 i_{2}(k) & = & \frac{J_0(k)}{2}-4 \int_0^1 \frac{r^5(1-r^2)}{(1+r^2)^3} J_0(k r) \ud r \\
 i_{3}(k) & = & \int_0^1 \frac{r^4}{(1+r^2)^2} J_1(k r) \ud r \\
 U_{22}(h) & = & \int_0^\infty e^{-k h} (i_{0}(k) i_{2}(k) + 8 (i_{3}(k))^2) \ud k \\
 i_{4}(k) & = & J_0(k) + 16 \int_0^1 \frac{r^5}{(1+r^2)^3} J_0(k r) r \ud r \\
 U_{40}(h) & = & \int_0^\infty e^{-k h} (64 (i_{1}(k))^2 + 4 i_{0}(k) i_{4}(k)) \ud k \\
 i_{5}(k) & = & - 6 J_0(k) + k J_1(k) + \nonumber \\
  & & 32 \int_0^1 \frac{r^7(2 - r^2)}{(1+r^2)^4} J_0(k r) \ud r \\
 i_{6}(k) & = & J_1(k) - 4 \int_0^1 \frac{r^4(1 - 3 r^2)}{(1+r^2)^3} J_1(k r) \ud r \\
 i_{7}(k) & = & 8 i_2(k) i_1(k) - i_5(k) i_0(k) + 8 i_3(k) i_6(k) \\
 U_{42}(h) & = & \int_0^\infty e^{-k h} i_7(k) \ud k \\
 i_{8}(k) & = & (62-k^2) J_0(k) - 16 k J_1(k) + \nonumber \\
 & & 128 \int_0^1 \frac{r^9(1 - 4 r^2 - r^4)}{(1+r^2)^5} J_0(k r) \ud r \\
 i_{9}(k) & = & k J_0(k) + 6 J_1(k) - \nonumber \\
 & & 32 \int_0^1 \frac{r^8(2 - r^2)}{(1+r^2)^4} J_1(k r) \ud r \\
 i_{10}(k) & = & \int_0^1 \frac{r^7}{(1+r^2)^3} J_2(k r) \ud r \\
 i_{11}(k) & = & (J_2(k))^2 + 16 i_3(k) i_9(k) +  i_0(k) i_8(k) + \nonumber \\
 & & 8 (i_2(k))^2 + 32 i_{10}(k) (J_2(k) + 8 i_{10}(k))\\
 U_{44}(h) & = & \frac{1}{32}\int_0^\infty e^{-k h} i_{11}(k) \ud k
\end{eqnarray}
These expressions, while compact, are not easy to evaluate numerically, 
because of the $k$ integration over infinite region of oscillating
products of Bessel's functions. This difficulty is especially pronounced
when evaluating $U_i(0)$, entering the magnetostatic functions $W_i(h)$.
To alleviate this difficulty the actual numerical formulas
in the {\tt MATHEMATICA} file {\tt frequency.m} are
further rearranged by using the following identity III 6.612(3) in 
Ref.~\onlinecite{Gradshtejn_Ryzhik}
\begin{eqnarray}
 \int_0^\infty & e^{- k h} & J_n (k r_1) J_n(k r_2) \ud k = \nonumber \\
 & & \frac{Q_{n-1/2}((h^2+r_1^2+r_2^2)/(2 r_1 r_2))}{\pi\sqrt{r_1 r_2}},
\end{eqnarray}
where $Q_m(x)$ is the Legendre function of the second kind. 
Some of the remaining  integrals were taken analytically. 
The result of this convergence improvement was checked against
the original expressions above.

\section{Magnetostatic functions for volume charges}
The expansion of the energy of the volume charges in powers 
of $|A|$ can be factored using the Lipshitz integral and Bessel's summation
theorem similarly to the energy of face charges. The integration across $Z$,
can be carried out analytically
\begin{eqnarray}
 & & \frac{1}{2g}\int_0^g\int_0^g e^{-k |z_1-z_2|} \ud z_1 \ud z_2 =
 \frac{f_\mathrm{MS}(k g)}{k} \\
 & & f_\mathrm{MS}(x) = 1 - \frac{1-e^{-x}}{x} .
\end{eqnarray}
Taking the remaining angular and radial integrals produces
\begin{equation}
 \label{eq:evolumeapp}
 e_\mathrm{v} =  |A|^2 V_2(r_V, g) + |a|^4 V_4(r_V, g)
\end{equation}
with the following magnetostatic functions
\begin{eqnarray}
 j_0(r_V,k) & = & 4 r_V^3 \int_0^{r_V} \frac{r^2}{(r^2+r_V^2)^2} J_1(k r) \ud r + \nonumber \\
 & &   \int_{r_V}^{1} r J_1(k) \ud k \\
 V_2(r_V,g) & = & 2 \int_0^\infty \frac{f_\mathrm{MG}(k g)}{k} (j_0(r_V, k))^2 \ud k \\
 j_1(r_V,k) & = & -J_2(k) + \int_{r_V}^{1} r^2 J_2( k r ) \ud r \\
 j_2(r_V,k) & = & k J_0(k) - 5 J_1(k) + r_V^4 J_1(k r_V) + \nonumber \\
 & & \int_{r_V}^1 r (16 - r^2) J_1(k r) \ud r + \nonumber \\
 & & 32 r_V^3 \int_0^{r_V} 
 \frac{r^4(r^4 +8 r_V^2 + 2 r^2 (4 - r_V^2)) J_1(k r)}
      {(r^2 + r_V^2)^4} \ud r \\
 j_3 (r_V,k) & = & \int_0^{r_V} \frac{r^5 J_2(k r)}{(r^2 + r_V^2)^3} \ud r \\
 j_4 (r_V,k) & = & (j_1(r_V,k))^2 +j_0(r_V,k)j_2(r_V,k)+ \nonumber \\
 & & 32 r_V^3 j_1(r_V,k) j_3(r_V,k) + 256 r_V^6 (j_3(r_V,k))^2 \\
 V_4(r_V,g) & = & \frac{1}{2} \int_0^\infty \frac{f_\mathrm{MG}(k g)}{k} j_4(r_V, k) \ud k .
 \end{eqnarray}
 
}{%
}
%
\end{document}